\documentclass[letterpaper]{article} % DO NOT CHANGE THIS
\usepackage{aaai2026}  % DO NOT CHANGE THIS
\nocopyright
\usepackage{times}  % DO NOT CHANGE THIS
\usepackage{helvet}  % DO NOT CHANGE THIS
\usepackage{courier}  % DO NOT CHANGE THIS
\usepackage[hyphens]{url}  % DO NOT CHANGE THIS
\usepackage{graphicx} % DO NOT CHANGE THIS
\usepackage{natbib}  % DO NOT CHANGE THIS AND DO NOT ADD ANY OPTIONS TO IT
\usepackage{caption} % DO NOT CHANGE THIS AND DO NOT ADD ANY OPTIONS TO IT
\usepackage{algorithm}
\usepackage{algorithmic}

\usepackage{newfloat}
\usepackage{listings}
\DeclareCaptionStyle{ruled}{labelfont=normalfont,labelsep=colon,strut=off} % DO NOT CHANGE THIS
\floatstyle{ruled}
\newfloat{listing}{tb}{lst}{}
\floatname{listing}{Listing}
\usepackage{multirow}

\title{Enabling Multi-Agent Systems as Learning Designers: Applying Learning Sciences to AI Instructional Design}
\author{
    Jiayi Wang\textsuperscript{\rm 1},
    Ruiwei Xiao\textsuperscript{\rm 2},
    Xinying Hou\textsuperscript{\rm 3},
    John Stamper\textsuperscript{\rm 2}
}
\affiliations{
    \textsuperscript{\rm 1}Northwestern University\\
    \textsuperscript{\rm 2}Carnegie Mellon University\\
    \textsuperscript{\rm 3}University of Michigan\\
    jiayiwang2025@u.northwestern.edu, ruiweix@andrew.cmu.edu, xyhou@umich.edu, jstamper@cmu.edu

}

\begin{document}

\maketitle

\begin{abstract}
    K-12 educators are increasingly using Large Language Models (LLMs) to create instructional materials. These systems excel at producing fluent, coherent content, but often lack support for high-quality teaching. The reason is twofold: first, commercial LLMs, such as ChatGPT and Gemini which are among the most widely accessible to teachers, do not come preloaded with the depth of pedagogical theory needed to design truly effective activities; second, although sophisticated prompt engineering can bridge this gap, most teachers lack the time or expertise and find it difficult to encode such pedagogical nuance into their requests. This study shifts pedagogical expertise from the user's prompt to the LLM's internal architecture. We embed the well-established Knowledge–Learning–Instruction (KLI) framework into a Multi-Agent System (MAS) to act as a sophisticated instructional designer. We tested three systems for generating secondary Math and Science learning activities: a Single-Agent baseline simulating typical teacher prompts; a role-based MAS where agents work sequentially; and a collaborative MAS–CMD where agents co-construct activities through conquer and merge discussion. The generated materials were evaluated by 20 practicing teachers and a complementary LLM-as-a-judge system using the Quality Matters (QM) K-12 standards. While the rubric scores showed only small, often statistically insignificant differences between the systems, the qualitative feedback from educators painted a clear and compelling picture. Teachers strongly preferred the activities from the collaborative MAS-CMD, describing them as significantly more creative, contextually relevant, and classroom-ready. Our findings show that embedding pedagogical principles into LLM systems offers a scalable path for creating high-quality educational content.
\end{abstract}


\section{Introduction}

The emergence of powerful Large Language Models (LLMs) since 2022 has rapidly transformed K–12 education, with surveys showing growing adoption for tasks like lesson planning \cite{Klopfer2024Generative, TAN2025100355}. Yet a persistent “prompting gap” limits their effectiveness: while teachers can improve prompt quality with practice, most lack the time or expertise to consistently generate high-quality outputs \cite{octavio2024chatgpt}. Without training or advanced tools, LLMs remain difficult to use as true instructional partners.

To address this gap, AI must evolve from passive text generators to active pedagogical partners by embedding learning science principles directly into content generation. We adopt the Knowledge–Learning–Instruction (KLI) framework, which connects knowledge components, learning processes, and instructional principles to support robust learning \cite{koedinger2012knowledge}.

To simulate the complex, multi-step reasoning process in the KLI framework, we employ Multi-Agent Systems (MAS), with core advantages “in its distributed decision-making and problem-solving capabilities" \cite{li2024survey}. This system is well-suited to the complex process of instructional design, allowing different agents to assume specific pedagogical roles and complete sub-tasks.

This paper investigates the following research question: Compared to a baseline of simple prompts teachers might use, do multi-agent systems guided by the KLI framework produce higher-quality learning activities? Our contributions are threefold: (1) we design and implement three distinct generative systems (SAS, MAS-Roles, and MAS-CMD) for learning activity generation; (2) we evaluate the outputs through a mixed-methods analysis involving practicing secondary teachers and a theory-grounded LLM-as-a-judge \cite{li2024llmsasjudgescomprehensivesurveyllmbased}; (3) we provide a quantitative analysis of not only the pedagogical quality of the outputs but also their computational cost, revealing a critical trade-off between efficiency and quality. This analysis goes beyond simply identifying the “best" system to understanding the practical costs associated with embedding deep pedagogical theory into generative AI, a crucial consideration for the development of scalable and accessible tools for educators.
\section{Related Work}

\subsection{Generative AI and MAS in Instructional Design}

Since late 2022, K–12 teachers have increasingly adopted LLMs to streamline lesson planning, activity design, and content differentiation. In fall 2023, 18\% of teachers reported using AI for teaching (with another 15\% having tried it) \cite{Diliberti2024K12AI}. By spring 2025, a study of 2232 public-school teachers found that 60\% had used AI during the 2024–25 school year, with 32\% using it at least weekly; common uses included preparing to teach (37\%), making worksheets/activities (33\%), and modifying materials to meet student needs (28\%) \cite{Ash2025GallupAI}. Weekly AI users save about 5.9 hours per week, nearly six weeks over a school year. Most teachers report AI improves their work quality \cite{Ash2025GallupAI}. Beyond single-agent prompting, multi-agent systems (MAS) show promise in K–12 instructional design by coordinating specialized agents to improve lessons, worksheets, and assessments. For instance, EduPlanner uses evaluator, optimizer, and question-analysis agents to create math lessons tailored to student knowledge \cite{zhang2025eduplannerllmbasedmultiagentsystems}, while the FACET framework simulates diverse learner profiles to generate individualized worksheets judged by teachers as well-structured and appropriate \cite{gonnermannmüller2025facetteachercentredllmbasedmultiagentsystemstowards}. A study shows that generator–critic MAS pipelines can yield high-quality AI literacy MCQs aligned with Bloom's Taxonomy and usable in classrooms \cite{wang2025generating}. These systems suggest that MAS can boost both scalability and pedagogical quality, providing richer, context-aware materials than single-agent prompting. Yet current MAS designs are not explicitly guided by learning sciences principles.

Studies caution that novice prompting often produces shallow, misaligned, or inconsistent materials. In mathematics, analyses of ChatGPT-generated tasks found tendencies toward procedural items, gaps in conceptual representations, and occasional inaccuracies, even after prompt variations, suggesting that unguided prompts can underdeliver on cognitive demand \cite{sapkota2024assessing}. An evaluation of AI-generated curriculum materials characterized some outputs as “a mile high and an inch deep,” with repetitive activities misaligned to specifications \cite{sawyer2025mile}. In foreign-language lesson planning, researchers found high variability across runs and historical–pedagogical biases that can surface without careful prompt design \cite{dornburg2024extentchatgptusefullanguage}. More broadly, studies show that without explicit scaffolds (role constraints, exemplars, rubrics), LLMs can produce hallucinations or surface-level plans that require significant teacher editing \cite{powell2024opportunities}. While more schools and districts are providing trainings to teachers about generative AI \cite{Diliberti2024K12AI}, simple and zero-shot prompting remains common \cite{chen2024systematic}, which may be insufficient for generating pedagogically robust materials.

\subsection{The Knowledge-Learning-Instruction (KLI) Framework}

Developed to bridge the cognitive science research and real-world educational practice, the KLI framework provides a structured, theory-driven approach to instructional design \cite{koedinger2012knowledge}. It states that effective instruction requires careful alignment among three core elements: Knowledge Components (KCs), which are the goals of learning (e.g., concepts, skills, facts); Learning Processes, which describe how learning occurs (e.g., memory and fluency building, induction and refinement, understanding and sense-making); and Instructional Principles, which are the specific methods used to facilitate those learning processes (e.g., spacing and testing, worked examples, prompted self-explanation) \cite{koedinger2012knowledge}. By using the KLI framework to guide the reasoning of our multi-agent systems, we are explicitly grounding our AI systems' design in a well-established, evidence-based instructional theory that aims to produce robust student learning. This helps us create a more principled and pedagogically-informed instructional design process. Recent work demonstrates how to apply learning sciences principles to generative AI, using the KLI framework as a guide to map knowledge types like debugging to specific tutoring protocols to elicit more effective, learning-oriented responses from the model \cite{xiao2025improvingstudentaiinteractionpedagogical}.

\subsection{LLM-as-a-Judge for Educational Evaluation}

The LLM-as-a-Judge paradigm refers to the use of a large language model to assess and score outputs, a technique being applied across various domains, including education \cite{li2024llmsasjudgescomprehensivesurveyllmbased}. This methodology comes with challenges. LLMs are known to exhibit systematic biases, including a preference for longer outputs, a tendency to favor the first option in a comparison, and a potential preference for outputs generated by models from the same developer family \cite{gu2025surveyllmasajudge}. When carefully designed, however, this approach offers significant advantages in scalability and speed over traditional human evaluation, and studies have shown that LLM judges can achieve high levels of agreement with human raters. In education-specific contexts, GPT-4-as-judge has shown moderate agreement with human raters when scoring feedback quality with analytic rubrics, though with a positivity bias that requires prompt and aggregation controls \cite{koutcheme2024open}. With the benefits and limitations of LLM-as-a-judge in mind, our study uses this method as a complement for expert evaluation.
\section{Methods}

% \subsection{Task Design}
% [Survey]
% [LO inclusion and exclusion criteria]

% randomly sampled 10 representative LOs from around 300 LOs in math and science national standards (x\% of the total), run it for 1 time, and let expert evaluate learning activities generated under each condition for one learning objective

% \begin{table}[ht]
% \centering
% \caption{Experimental setup for expert evaluation of generated learning activities}
% \label{tab:setup}
% \begin{tabular}{lcccccc}
% \toprule
% \textbf{Setup} & \textbf{LOs} & \textbf{\% of National Standard} & \textbf{\# of Variance Runs} & \textbf{\# of Graders Per LO} & \textbf{\# of TASKS PER Graders} \\
% \hline
% \midrule
% #1 (current) & 10 & $x$\% & 1 & 10 & 3 \\
% #2 & 10 & $x$\% & 5 & 2 & 3 \\
% #3 & 50 & $x$\% & 1 & 2 & 3 \\
% \hline
% \bottomrule
% \end{tabular}
% \end{table}

\subsection{System Design and Implementation}

The core of this study is designing, implementing, and comparing three systems for generating K–12 learning activities

\subsubsection{Baseline Single-Agent System (SAS)}
The Single-Agent System (SAS) was designed as a baseline to represent a naive, non-expert interaction with an LLM. Prior research studies show that non-experts in prompt engineering within educational contexts often rely on simple requests, often consisting of a single verb, and use basic “copy-paste" strategies to provide contextual information \cite{octavio2024chatgpt,tassoti2024assessment}. This SAS design able the measurement of the “value-added" by more sophisticated, theory-informed architectures. Its prompts are constructed by appending contextual information: the subject domain, grade level, standard alignment, and the learning objective, to the phrase “Generate learning activity". 

\subsubsection{Role-Based Multi-Agent System (MAS-Roles)}
The Role-Based Multi-Agent System (MAS-Roles) is guided by the KLI theoretical framework. Our system operationalizes this theory through a sequential pipeline of specialized agents, representing a centralized, task-decomposition multi-agent architecture where a complex problem is broken down into manageable subtasks. This system does not merely use concepts from the KLI framework as input, but embodies the analytical process of the framework in its very structure. The system functions as a five-stage automated and iterative pipeline, where the output of each agent serves as the primary input for the subsequent agent. {\bf KC Agent:} Receives the subject domain, grade level, standard alignment, and learning objective and identifies the core knowledge components (KCs), such as facts, concepts, principles, or procedures. {\bf Learning Process Agent:} Analyzes the KCs to determine the most appropriate learning process. {\bf Instructional Principle Agent:} Based on the selected learning process, this agent chooses a relevant instructional principle. {\bf Design Agent:} Synthesizes the outputs from the previous three agents to generate a complete learning activity. {\bf Feedback Agent:} Checks the generated activity for coherence and alignment with the initial contextual information, and decides if more iterations and modifications are needed.

\subsubsection{Multi-Agent System with Conquer and Merge Discussion (MAS-CMD)}
The Multi-Agent System with Conquer and Merge Discussion (MAS-CMD) \cite{wang2024rethinkingboundsllmreasoning} implements a more dynamic, collaborative architecture designed to simulate a professional discussion. {\bf Initial Generation:} Three distinct agents are instantiated. Each is given one of the five “teacher personas": Behaviorist, Constructivist, Aesthetic, Ecological, Integrated Social-Emotional \cite{mcconnell2020lesson}, for curriculum design to encourage diverse outputs. All three agents receive the same KLI framework guidance as the MAS-Roles system and independently create a draft learning activity. {\bf Collaborative Discussion:} The three agents engage in a structured, multi-turn dialogue. Each agent presents its draft and asks the other two agents for feedback, then revises its draft based on that feedback. {\bf Final Selection:} A final decision agent reviews the entire discussion transcript and the three revised drafts. It then chooses the most appropriate one and outputs the final learning activity.

\subsection{Experimental Design and Procedure}
A corpus of 10 learning objectives, 5 from Math and 5 from Science, was selected from the human-curated lesson plan library of the Oak Ridge Institute for Science and Education (ORISE) \cite{ORISELessonPlans}. ORISE is a U.S. Department of Energy institute, providing a repository of high-quality, federally supported STEM resources for educators. The learning objectives selected from this source ensures that the generation tasks are grounded in authentic, curriculum-aligned educational goals. These objectives include subject domain, grade level, and standard alignment information. 

Each of the three systems (SAS, MAS-Roles, MAS-CMD) was used with the gemini-2.5-flash model to generate a learning activity for each of the 10 objectives, resulting in a total of 30 unique learning activities for expert human evaluation. In addition, an extra 30 learning activities were generated using each of the following models: gemini-2.5-pro, gemini-2.0-flash, and gemini-2.0-flash-lite, yielding 90 additional learning activities. Combined with the 30 generated using gemini-2.5-flash, all 120 learning activities were evaluated using the LLM-as-a-Judge methodology. The models were selected to represent a range of model capabilities. Default temperature and top-p settings were used throughout the generation process to encourage variation in the outputs.

\subsubsection{Participants}
We recruited 21 US-based secondary math (N=12) and science (N=9) teachers via Prolific to serve as domain experts in evaluating the quality of the generated learning activities under all three conditions. Pre-screening criteria required participants to be located in the United States and teach grades 7–12 mathematics or science, thereby ensuring relevant subject-matter expertise. Each participant received \$4–6 in compensation for completing all required tasks, which took approximately 15–30 minutes. 

\subsubsection{Evaluation Rubric}
To ensure a rigorous and standardized evaluation, we adapted the Quality Matters (QM) K–12 Rubric (Table~\ref{tab:qm_rubric}), focusing on four review standards from General Standard 5 (Learning Activities and Learner Interaction) \cite{QM2025K12Standards6e}. Rather than asking evaluators to just assign numbers, we presented multiple-choice, plain-language descriptors that correspond to the rubric's performance levels. Each item was aligned to the 0–3 scale used in the QM rubric (0–2 for 5.4C), and evaluators selected the descriptor that best matched the activity; we then coded responses to the associated numeric level for analysis. For example, for Standard 5.1C, the options were framed as: clearly aligned to objectives and standards and directly supportive of achieving them (3 points), generally aligned but with limited or inconsistent support (2 points), alignment is weak or unclear (1 point), and does not support the stated objectives or standards (0 point). Similar descriptions were provided for the other criteria. This rubric was used in both human and LLM evaluation.

\begin{table}[ht]
    \centering
    \fontsize{9pt}{11pt}\selectfont
    \begin{tabular}{p{0.7\columnwidth} p{0.2\columnwidth}}
        % \hline
        \textbf{Criteria} & \textbf{Score Range} \\
        % \hline
        \textbf{5.1 C} The learning activities help learners achieve the stated learning objectives and content standards. & 0--3\\
        \textbf{5.2 C} Learning activities provide opportunities for interactions that support active learning. & 0--3 \\
        \textbf{5.3 C} The expectations for instructor responsiveness and feedback are clearly stated. (This was adapted to assess how well the generated activity plan outlined these expectations). & 0--3\\
        \textbf{5.4 C} The requirements for learner interaction are clearly stated. & 0--2\\
        % \hline
    \end{tabular}
    \caption{Quality Matters Evaluation Rubric}
    \label{tab:qm_rubric}
\end{table}

To enable a transparent and reliable LLM-as-a-judge process, we created the Integrated Learning Sciences Evaluation Rubric (Table~\ref{tab:ils_rubric}), which synthesizes classic instructional design and learning sciences. The Foundational Alignment \& Clarity domain draws on Constructive Alignment and Backward Design to judge coherence among outcomes, activities, and assessments \cite{biggs2022teaching,wiggins2005understanding}. Cognitive Complexity \& Rigor operationalizes the revised Bloom's Taxonomy to locate the primary cognitive demand \cite{anderson2001taxonomy}. Instructional Scaffolding \& Process is grounded in Gagné's Nine Events of Instruction, emphasizing activation of prior knowledge, clear guidance, structured practice, and feedback \cite{gagne2005principles}. Learner Engagement \& Motivation and Inclsivity \& Accessibility are guided by Universal Design for Learning (UDL), providing multiple means of engagement, representation, and action/expression, as well as motivational supports \cite{cast_udl_guidelines}. Together, these five domains yield a 17-criterion rubric that evaluates the quality of learning activities.

\begin{table}[ht]
    \centering
    \fontsize{9pt}{11pt}\selectfont
    \begin{tabular}{p{0.25\columnwidth}|p{0.65\columnwidth}}
    % \hline
    \textbf{Dimension} & \textbf{Criteria} \\
    \hline
    
    \multirow{4}{=}{I. Foundational Alignment \& Clarity} 
    & 1.1 Clarity of Learning Outcome \\
    & 1.2 Activity-Outcome Alignment \\
    & 1.3 Assessment-Outcome Alignment \\
    & 1.4 Learner Context Awareness \\
    \hline
    
    \multirow{3}{=}{II. Cognitive Complexity \& Rigor} 
    & 2.1 Cognitive Level (Bloom's) \\
    & 2.2 Promotion of Higher-Order Thinking \\
    & 2.3 Authenticity and Application \\
    \hline
    
    \multirow{4}{=}{III. Instructional Scaffolding \& Process} 
    & 3.1 Activation of Prior Knowledge \\
    & 3.2 Clear Presentation \& Guidance \\
    & 3.3 Opportunity for Practice \& Performance \\
    & 3.4 Provision of Feedback \\
    \hline
    
    \multirow{3}{=}{IV. Learner Engagement \& Motivation} 
    & 4.1 Recruiting Interest \\
    & 4.2 Sustaining Effort \& Persistence \\
    & 4.3 Fostering Self-Regulation \\
    \hline
    
    \multirow{3}{=}{V. Inclusivity \& Accessibility} 
    & 5.1 Multiple Means of Representation \\
    & 5.2 Multiple Means of Action \& Expression \\
    & 5.3 Removal of Barriers \\
    % \hline
    
    \end{tabular}
    \caption{Integrated Learning Sciences Evaluation Rubric}
    \label{tab:ils_rubric}
    \end{table}

\subsection{Data Analysis}

We analyzed the data using a mixed-methods approach. For human expert ratings, each rater evaluated all three systems (SAS, MAS-Roles, MAS-CMD) on QM criteria 5.1 C–5.4 C and a total score. Within-subject differences were tested using repeated-measures ANOVAs (factor: System), and when the omnibus test was significant ($\alpha = .05$), Holm-adjusted paired $t$-tests were conducted. Non-parametric tests (Friedman and Wilcoxon) were used as robustness checks. Inter-rater consistency was quantified using Fleiss’ $\kappa$. For LLM-based automatic evaluations on the QM and Integrated Learning Sciences rubrics, we report descriptive statistics (means and standard deviations) only. Operational efficiency metrics (time, tokens, requests) and self-reported AI tool usage were also summarized descriptively to contextualize the results. All quantitative analyses were performed using Python. For the qualitative analysis, written feedback from the K–12 instructors was analyzed using thematic analysis \cite{braun2006using} to identify recurring patterns and themes related to the perceived strengths and weaknesses of the activities generated by each system.

\section{Results}

One participant's response was excluded from analysis due to providing off-topic written feedback and completing the task in an unusually short time, indicating insufficient engagement. For evaluation, we relied on data from the remaining 20 participants (11 Math, 9 Science).

\begin{figure}[t]
    \centering
    \includegraphics[width=1\columnwidth]{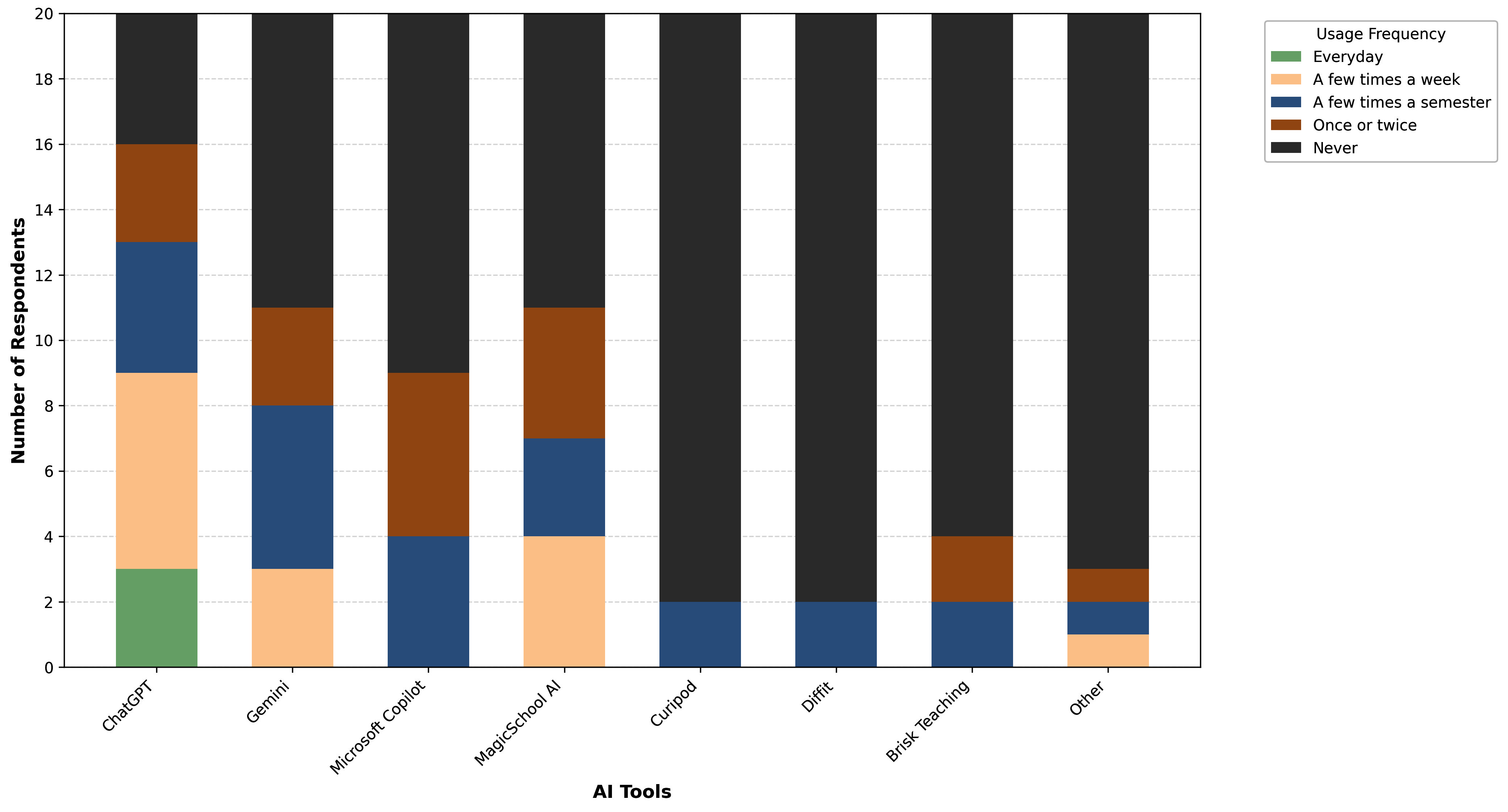}
    \caption{AI Tools Usage Frequency}
    \label{fig:ai_tools}
\end{figure}

Self-reported AI usage patterns (Fig.~\ref{fig:ai_tools}) indicate that familiarity is concentrated in general-purpose chatbots. Several teachers reported using ChatGPT (and to a lesser extent Gemini) weekly or daily, whereas most domain-specific tools (e.g., Curipod, Diffit, Brisk Teaching) were rarely or never used. This distribution contextualizes the human ratings: evaluators were more accustomed to general chatbots than to specialized K–12 authoring tools.

Computational costs vary across system designs. (Table~\ref{tab:la_stats}). SAS was fastest and lightest (25$\pm$16 s; $\sim$3.4k tokens; one request). MAS-Roles required roughly $3\times$ the time and $\sim$7$\times$ the tokens, while MAS-CMD traded the highest rubric scores for substantially greater latency and usage (272$\pm$168 s; $\sim$71.6k tokens; 13 requests). These results quantify a quality–efficiency trade-off across design strategies.

\begin{table}[ht]
\centering
\fontsize{9pt}{11pt}\selectfont
\begin{tabular}{lcccc}
\hline
System & Time (sec) & Tokens Used & Requests Made \\
\hline
SAS & 25±16 & 3,376±1,601 & 1.00±0.00 \\
MAS-Roles & 77±45 & 22,497±7,157 & 5.42±1.13 \\
MAS-CMD & 272±168 & 71,638±25,222 & 13.00±0.00 \\
\hline
\end{tabular}
\caption{Learning Activity Generation Statistics by System}
\label{tab:la_stats}
\end{table}

% \subsection{Quantitative Analysis of Human Evaluation}
% \begin{table}[h]
%     \centering
%     \begin{tabular}{lrrrrr}
%     \hline
%               & 5.1 C & 5.2 C & 5.3 C & 5.4 C & Total \\
%     \hline
%     SAS       & 2.70 & 2.75 & 2.35 & 1.75 & 9.55 \\
%     MAS-Roles & 2.60 & 2.55 & 2.20 & 1.70 & 9.05 \\
%     MAS-CMD   & 2.85 & 2.95 & 2.45 & 1.80 & 10.05 \\
%     \hline
%     \end{tabular}
%     \caption{QM Rubric Expert Evaluation Average Score}
%     \label{tab:hu_score}
% \end{table}

\subsection{Human Evaluation}

\subsubsection{MAS-CMD Received Highest Expert Scores, Statistically Significant on Criterion 5.2 C Active Learning}

ANOVA was conducted to compare expert evaluations of the three instructional systems across four criteria (5.1 C–5.4 C) and a composite total score. Each of the twenty experts provided ratings for all three systems on each criterion, enabling within-subject comparisons.

Descriptive analyses indicated that MAS-CMD consistently received the highest average ratings across all four criteria and the total score, whereas MAS-Roles tended to receive the lowest scores. For example, the mean total score was $M = 10.05$, 95\% CI [9.42, 10.68] for MAS-CMD, compared with $M = 9.55$, 95\% CI [8.92, 10.18] for SAS, and $M = 9.05$, 95\% CI [8.28, 9.82] for MAS-Roles. These were computed using the standard confidence interval formula:

\[
\bar{x} \pm t_{\alpha/2,\, df} \cdot \frac{s}{\sqrt{n}}
\]

ANOVA revealed a statistically significant effect of system on criterion~5.2, $F(2, 38) = 4.75$, $p = .014$, $\eta_p^2 = .20$, indicating a large effect size. Post-hoc pairwise comparisons with Holm correction showed that MAS-CMD was rated significantly higher than MAS-Roles ($p = .050$), with no other significant pairwise differences. For criteria 5.1, 5.3, and 5.4, the omnibus tests were non-significant (all $p > .20$). For the total score, the omnibus ANOVA did not reach conventional significance, $F(2, 38) = 2.66$, $p = .083$, although the non-parametric Friedman test suggested an overall difference across systems, $\chi^2(2) = 6.24$, $p = .044$. This statistic was computed using the Friedman test formula:
\[
\chi^2_F = \frac{12}{n k (k+1)} \sum_{j=1}^k R_j^2 - 3n(k+1)
\]
where \(n\) is the number of raters, \(k\) is the number of systems, and \(R_j\) is the total rank for system \(j\). However, post-hoc Wilcoxon signed-rank tests with Holm correction did not reveal any pairwise comparisons that were statistically significant.

To assess the consistency of expert judgments, we computed Fleiss’ $\kappa$ across the twenty raters for each criterion and for the combined scores. Agreement was low overall: criterion~5.1 yielded $\kappa = 0.047$ (slight agreement), criterion~5.2 $\kappa = 0.053$ (slight), criterion~5.3 $\kappa = -0.030$ (poor), and criterion~5.4 $\kappa = -0.043$ (poor). When all criteria were aggregated, the overall $\kappa$ was $-0.016$, indicating agreement below chance level. These results demonstrate that expert evaluations were highly inconsistent, limiting the reliability of the observed effects.

% Taken together, these results suggest that MAS-CMD is generally rated most favorably by experts, with the strongest evidence for superiority emerging on criterion~5.2. Although the overall pattern consistently favored MAS-CMD, differences across systems were small in magnitude for most criteria and did not consistently reach statistical significance after correction for multiple comparisons.

\begin{figure}[t]
    \centering
    \includegraphics[width=1\columnwidth]{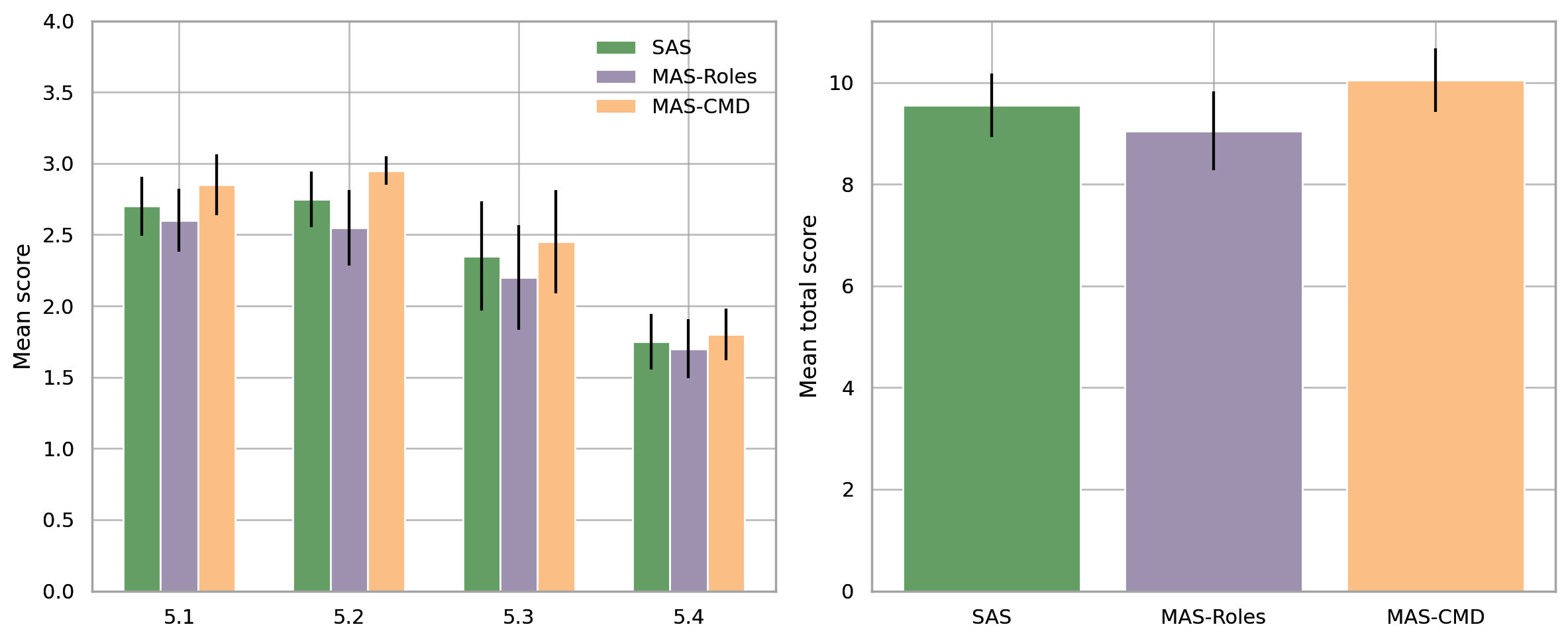} 
    \caption{Ratings by System}
    \label{fig:ratings_by_system}
\end{figure}

\subsubsection{Teachers Found MAS-CMD Most Creative and Complete; SAS Too Basic, MAS-Roles Too Generic}

The SAS produced relevant lesson plans but often lacked the polish, completeness, and innovation that teachers desire. Educators praised the SAS for generating relevant content, such as a lesson on screen time, and for creating engaging activities that positioned students as “detectives". Some of its outputs were considered “a really solid lesson plan" that could serve as a “springboard" for a teacher's own ideas. However, SAS was also criticized for a lack of innovation and for failing to provide complete, classroom-ready packages. Teachers noted that student handouts “would need a lot of editing" and expressed a desire for a “print-ready worksheet". Other common issues included pedagogical and structural flaws, such as a “confusing" warm-up activity, poor pacing where the “timing seems really far off," and lesson designs that felt too “rigid" for the classroom.

The MAS-Roles system was often seen as a step up from the SAS, producing well-structured lessons, but was sometimes perceived as generic or unresponsive to specific requests. The system was frequently commended for its strong pedagogical design, which included a “step by step approach" and started with “investigation instead of direct instruction". Teachers found the concepts, like “Geometry Detectives," to be fun and engaging, and they appreciated the inclusion of both formative and summative assessments within the lesson plans. Despite these strengths, feedback indicated the MAS-Roles system could feel “generic and impersonal," with one teacher remarking that a lesson reminded them of existing curricula. A significant point of failure was its occasional inability to follow instructions, such as a request for differentiation that was absent from the final plan. Some of the generated activities were also seen as having “very minimal interaction between students," limiting their collaborative potential.

The MAS-CMD system received the most enthusiastic praise, consistently lauded for its creativity, real-world context, and comprehensive, teacher-friendly resources. Creativity was a standout feature of the MAS-CMD, with teachers describing its ideas as “fantastic" and “so creative". The system excelled at creating “a strong real-world context," such as an “urban planning theme" that made geometry “meaningful" for students. It was also highly valued for providing a complete package for educators, including a “worksheet and exit ticket" which one teacher called “Super helpful," and “teacher dialogue" that was deemed “very helpful" for new teachers. The primary critiques of the MAS-CMD were practical in nature. Some lessons were described as being “a lot to unpack" and having potential “pacing issues," suggesting they might need to be “broken up into some days" to be implemented effectively. Other minor feedback mentioned the need for more explicit guidance on the use of classroom tools or assessment rubrics.

Several themes emerged across feedback, highlighting what educators value in AI-generated materials. A primary theme was the importance of student engagement and active learning, as all systems were praised when they produced activities that involved students through relevant topics, investigation, and movement. Another crucial theme was teacher utility and completeness. The most valuable outputs were those that saved teachers time by providing a complete package with worksheets, assessments, and guidance. Educators also consistently evaluated the outputs based on pedagogical soundness, looking for logical flow, clear objectives, and differentiation, with the MAS outputs generally seen as more robust. Finally, the feedback showed a strong preference for meaningful context, where deep, thematic integration that positioned students in a role like “Polygon Architect" was seen as far more powerful than simply covering a topic.

\subsection{LLM-as-a-Judge}
\subsubsection{LLM Ratings Favored MAS-CMD When Activities Were Generated by Stronger Models}

\begin{table*}[h]
    \centering
    \fontsize{9pt}{11pt}\selectfont
    \begin{tabular}{|l|l|c|c|c|c|c|}
    \hline
    \textbf{Model} & \textbf{System} & \textbf{QM 5.1 C} & \textbf{QM 5.2 C} & \textbf{QM 5.3 C} & \textbf{QM 5.4 C} & \textbf{QM Total} \\
     & & \textbf{Alignment} & \textbf{Active Learning} & \textbf{Feedback} & \textbf{Interaction} & \textbf{Score} \\
    \hline
    \multirow{3}{*}{\begin{tabular}[c]{@{}l@{}}gemini-2.0-\\flash-lite\end{tabular}} 
    & SAS & 3.0 (0.0) & 3.0 (0.0) & 2.0 (0.0) & 2.0 (0.0) & 10.0 (0.0) \\
    \cline{2-7}
    & MAS-Roles & 3.0 (0.0) & 3.0 (0.0) & 2.5 (0.527) & 1.9 (0.316) & \textbf{10.4 (0.699)} \\
    \cline{2-7}
    & MAS-CMD & 3.0 (0.0) & 3.0 (0.0) & 2.0 (0.0) & 2.0 (0.0) & 10.0 (0.0) \\
    \hline
    \multirow{3}{*}{\begin{tabular}[c]{@{}l@{}}gemini-2.0-\\flash\end{tabular}} 
    & SAS & 3.0 (0.0) & 3.0 (0.0) & 2.2 (0.422) & 1.9 (0.316) & 10.1 (0.568) \\
    \cline{2-7}
    & MAS-Roles & 3.0 (0.0) & 3.0 (0.0) & 2.3 (0.483) & 2.0 (0.0) & \textbf{10.3 (0.483)} \\
    \cline{2-7}
    & MAS-CMD & 3.0 (0.0) & 3.0 (0.0) & 2.3 (0.483) & 1.9 (0.316) & 10.2 (0.632) \\
    \hline
    \multirow{3}{*}{\begin{tabular}[c]{@{}l@{}}gemini-2.5-\\flash\end{tabular}} 
    & SAS & 3.0 (0.0) & 3.0 (0.0) & 2.2 (0.422) & 2.0 (0.0) & 10.2 (0.422) \\
    \cline{2-7}
    & MAS-Roles & 3.0 (0.0) & 3.0 (0.0) & 2.2 (0.422) & 1.9 (0.316) & 10.1 (0.568) \\
    \cline{2-7}
    & MAS-CMD & 3.0 (0.0) & 3.0 (0.0) & 2.6 (0.516) & 2.0 (0.0) & \textbf{10.6 (0.516)} \\
    \hline
    \multirow{3}{*}{\begin{tabular}[c]{@{}l@{}}gemini-2.5-\\pro\end{tabular}} 
    & SAS & 3.0 (0.0) & 3.0 (0.0) & 2.0 (0.0) & 1.9 (0.316) & 9.9 (0.316) \\
    \cline{2-7}
    & MAS-Roles & 3.0 (0.0) & 3.0 (0.0) & 1.9 (0.738) & 2.0 (0.0) & 9.9 (0.738) \\
    \cline{2-7}
    & MAS-CMD & 3.0 (0.0) & 3.0 (0.0) & 2.0 (0.0) & 2.0 (0.0) & \textbf{10.0 (0.0)} \\
    \hline
    \end{tabular}
    \caption{QM Rubric LLM Evaluation by Model and System: Mean (Standard Deviation)}
    \label{tab:qm_performance}
    \end{table*}

Across automatic evaluations, absolute differences between systems were small but consistent. On the QM rubric (Table~\ref{tab:qm_performance}), all model–system pairs reached the ceiling on Alignment (5.1 C) and Active Learning (5.2 C), so variation came almost entirely from Feedback (5.3 C) and Interaction (5.4 C). The strongest total was observed for gemini-2.5-flash + MAS-CMD ($M=10.6$, SD $=0.52$), driven by a higher Feedback score ($M=2.6$). With smaller backbones, MAS-Roles was slightly favored (e.g., gemini-2.0-flash: $M=10.3$), whereas SAS rarely led. Importantly, the largest gap between systems within a model was $\leq 0.7$ points ($<7\%$ of scale), indicating incremental gains rather than large separations.

\begin{table*}[h]
    \centering
    \fontsize{9pt}{11pt}\selectfont
    \begin{tabular}{|l|l|c|c|c|c|c|c|}
    % \begin{tabular*}{\textwidth}{@{\extracolsep{\fill}}|l|l|c|c|c|c|c|@{}}
    \hline
    \textbf{Model} & \textbf{System} & \textbf{Dim 1 Sum} & \textbf{Dim 2 Sum} & \textbf{Dim 3 Sum} & \textbf{Dim 4 Sum} & \textbf{Dim 5 Sum} & \textbf{Total Sum} \\
     & & \textbf{Foundational} & \textbf{Cognitive} & \textbf{Scaffolding} & \textbf{Engagement} & \textbf{Inclusivity} & \textbf{Score} \\
    \hline\multirow{3}{*}{gemini-2.0-flash-lite}
    & SAS & 16.0 (0.000) & 11.1 (1.101) & 15.5 (0.527) & 11.5 (0.527) & 12.0 (0.000) & \textbf{66.1 (2.079)} \\
    \cline{2-8}
    & MAS-Roles & 16.0 (0.000) & 10.7 (0.949) & 15.7 (0.483) & 11.6 (0.966) & 11.9 (0.316) & 65.9 (2.331) \\
    \cline{2-8}
    & MAS-CMD & 16.0 (0.000) & 10.4 (0.966) & 15.4 (0.699) & 11.7 (0.483) & 12.0 (0.000) & 65.5 (1.841) \\
    \hline
    \multirow{3}{*}{gemini-2.0-flash}
    & SAS & 16.0 (0.000) & 10.8 (1.033) & 15.6 (0.516) & 11.4 (0.699) & 12.0 (0.000) & 65.8 (2.098) \\
    \cline{2-8}
    & MAS-Roles & 16.0 (0.000) & 11.3 (0.823) & 16.0 (0.000) & 12.0 (0.000) & 12.0 (0.000) & \textbf{67.3 (0.823)} \\
    \cline{2-8}
    & MAS-CMD & 16.0 (0.000) & 11.2 (0.789) & 15.8 (0.422) & 11.8 (0.422) & 12.0 (0.000) & 66.8 (1.549) \\
    \hline
    \multirow{3}{*}{gemini-2.5-flash}
    & SAS & 16.0 (0.000) & 11.2 (0.919) & 15.8 (0.422) & 11.7 (0.483) & 12.0 (0.000) & 66.7 (1.703) \\
    \cline{2-8}
    & MAS-Roles & 16.0 (0.000) & 11.5 (0.707) & 16.0 (0.000) & 12.0 (0.000) & 12.0 (0.000) & 67.5 (0.707) \\
    \cline{2-8}
    & MAS-CMD & 16.0 (0.000) & 11.8 (0.632) & 16.0 (0.000) & 12.0 (0.000) & 12.0 (0.000) & \textbf{67.8 (0.632)} \\
    \hline
    \multirow{3}{*}{gemini-2.5-pro}
    & SAS & 16.0 (0.000) & 11.1 (0.876) & 15.8 (0.422) & 11.7 (0.483) & 12.0 (0.000) & 66.6 (1.647) \\
    \cline{2-8}
    & MAS-Roles & 16.0 (0.000) & 11.2 (0.789) & 15.6 (0.516) & 11.6 (0.516) & 12.0 (0.000) & 66.4 (1.713) \\
    \cline{2-8}
    & MAS-CMD & 16.0 (0.000) & 11.7 (0.483) & 15.8 (0.422) & 11.8 (0.422) & 12.0 (0.000) & \textbf{67.3 (1.252)} \\
    \hline
    \end{tabular}
    \caption{Integrated Learning Sciences Rubric LLM Evaluation by Model and System: Mean (Standard Deviation)}
    \label{tab:judge_dimension_sums}
    \end{table*}

\subsubsection{MAS-CMD and MAS-Roles Performed Similarly on Learning Sciences Criteria}

On the Integrated Learning Sciences rubric (Table~\ref{tab:judge_dimension_sums}) show a similar pattern. Foundational and Inclusivity dimensions saturated (both near their maxima across conditions), while Cognitive and Engagement provided the main differentiation. With the higher-capacity model, gemini-2.5-flash + MAS-CMD achieved the highest total ($M=67.8$), MAS-Roles led on gemini-2.0-flash ($M=67.3$), and SAS was marginally best on gemini-2.0-flash-lite ($M=66.1$). Standard deviations were uniformly small ($\leq 0.96$ on subscales; $\leq 2.33$ on totals), suggesting low run-to-run variability for the LLM judge.

% Taken together, the automatic judges partially mirror the human pattern: MAS-CMD tends to perform best—especially on feedback-related criteria—when paired with stronger base models, while MAS-Roles remains competitive on mid-tier models. However, improvements are modest in magnitude, underscoring that all three orchestration strategies produce broadly similar rubric scores under automatic evaluation.

\section{Discussion}

This study's central finding is the notable divergence between the quantitative and qualitative evaluations of the AI-generated learning activities. While quantitative analysis based on the Quality Matters (QM) rubric yielded statistically weak results, qualitative feedback from practicing teachers revealed a clear preference: the collaborative multi-agent system guided by the KLI framework (MAS-CMD) was a significantly more effective and valuable tool for instructional design.

\subsection{Interpreting the Tension Between Quantitative and Qualitative Results}

A skeptical reviewer might point to the non-significant ANOVA results for the total score and the low inter-rater reliability (Fleiss' Kappa) as evidence that no meaningful difference exists between the systems. We argue, however, that this inconsistency is itself a meaningful finding. It suggests that standardized rubrics, even well-established ones like QM, may fail to capture the nuanced, multifaceted nature of what experienced educators perceive as “quality" in a lesson plan. The low agreement among experts highlights that instructional quality is not a monolithic concept but a subjective judgment influenced by individual teaching styles, classroom contexts, and pedagogical priorities. 

In this light, the rich, consistent themes from the qualitative analysis become paramount. Teachers did not just rate the MAS-CMD outputs higher; they articulated why. Their praise centered on specific, high-value attributes that the rubric may not fully weigh: creativity (“so creative," “fantastic idea"), deep contextual relevance (the “urban planning theme makes the geometry meaningful"), and, crucially, teacher-centric utility (the inclusion of a “worksheet and exit ticket" was “Super helpful"). This suggests that the true value of an AI tool for educators lies not just in its alignment with abstract standards but in its ability to produce innovative, engaging, and classroom-ready materials that reduce teacher workload. 

\subsection{The Complementary Role of LLM-as-a-Judge}

To complement the human evaluation, this study also employed an LLM-as-a-judge methodology, which offers scalability and speed but comes with its own set of considerations. The results from the LLM judge align more closely with the inconclusive human quantitative ratings than the qualitative feedback. Across both the QM and the Integrated Learning Sciences rubrics, the LLM judge found only small, incremental differences between the systems, with the largest gap being less than 7\% of the total scale. Furthermore, the LLM judge reached a “ceiling effect" on several criteria, such as Foundational Alignment and Inclusivity, where all systems scored near the maximum, limiting the rubric's ability to differentiate the outputs meaningfully. 

This outcome reinforces our interpretation that current standardized rubrics may be insufficient for capturing true instructional quality. The study is candid about the known limitations of the LLM-as-a-judge paradigm, including systematic biases like a preference for longer outputs or a “positivity bias". While the LLM judge provided a consistent and low-variance evaluation, it was ultimately less discerning than the expert teachers. The qualitative feedback identified clear and significant differences in creativity and utility that the automated scoring process, constrained by the rubric's structure, failed to detect. This suggests that while LLM-as-a-judge is a valuable tool for scalable analysis, it should not be a substitute for expert human judgment, especially when evaluating complex, creative artifacts like learning activities.

\subsection{The Significance of Multi-Agent Architecture}

The qualitative preference for MAS-CMD over MAS-Roles suggests that not all multi-agent systems are created equal. While the sequential, pipeline-based MAS-Roles produced solid lessons, it was sometimes perceived as “generic" and unresponsive. In contrast, the MAS-CMD's “conquer and merge discussion" architecture, where multiple agent “personas" generate, critique, and revise ideas, better simulates the dynamic, collaborative, and iterative nature of professional curriculum design. This collaborative process likely accounts for the greater creativity and coherence that teachers praised in its outputs. This finding contributes to the growing body of work on multi-agent systems, suggesting that architectures that foster emergent synthesis through collaboration may be particularly well-suited for complex, creative design tasks like instructional design.

\subsection{Limitations and Future Directions}

First, a crucial limitation is that this study evaluated the design of the learning activities, not their implementation. The activities were reviewed by teachers and LLMs but were not tested in real-world classrooms with students. Therefore, while the results speak to the perceived quality and potential of the AI-generated materials, they do not provide direct evidence of their impact on student engagement or learning outcomes. Second, the low inter-rater reliability on the quantitative measures, while interpretable, is a significant constraint on the generalizability of those specific findings. Future work should aim to develop or adapt evaluation rubrics that better capture the dimensions of quality (e.g., creativity, utility) that educators highlighted in their qualitative feedback. Third, our sample of 20 practicing teachers is relatively small and was limited to secondary math and science in the U.S.. Further research should explore a wider range of subjects, grade levels, and cultural contexts. Finally, this study was limited to a single pedagogical framework (KLI) and a specific set of MAS architectures. The results here should motivate future investigations into how other instructional design models and theories could be embedded into different agentic workflows.

\section{Conclusion}

This study examined whether multi-agent systems (MAS) guided by the KLI framework could generate higher-quality learning activities than a baseline single-agent system. Educators consistently preferred the KLI-guided systems, especially the collaborative MAS-CMD, which was praised for creativity, pedagogical integrity, and classroom readiness.

The key contribution is showing that embedding learning science principles directly into AI architectures is an effective way to produce pedagogically sound materials. This approach bridges the “prompting gap,” shifting AI from a content generator to a pedagogical partner. Findings also highlight that MAS architecture matters: a collaborative, discussion-based model outperformed a sequential one in producing innovative, coherent designs.

Although the most effective system was also the most computationally expensive, the qualitative value for teachers was clear. Future work should focus on optimizing these architectures and developing evaluation methods that reflect educators' needs and values. Overall, this study points to a promising direction for AI in education: LLM tools that are pedagogically intelligent by design.

\bibliography{aaai2026}

\end{document}